\documentclass[twocolumn,showpacs,preprintnumbers,amsmath,amssymb]{revtex4}
\input psfig.sty

\usepackage{graphicx}
\usepackage{dcolumn}
\usepackage{bm}

\begin{document}

\title{Simplified quartessence cosmology}

\author{J. A. S. Lima$^{1}$} \email{limajas@astro.iag.usp.br}
\author{J.
V. Cunha$^{1}$} \email{cunhajv@astro.iag.usp.br} \author{J. S.
Alcaniz$^{2}$}\email{alcaniz@on.br}

\affiliation{$^{1}$Instituto de
Astronomia, Geof\'{\i}sica e Ci\^encias Atmosf\'ericas, USP,
05508-900 S\~ao Paulo, SP, Brasil}

\affiliation{$^{2}$Departamento de Astronomia, Observat\'orio Nacional, 20921-400 Rio de Janeiro, RJ, Brasil}

\date{\today}

\begin{abstract}

We propose a new class of accelerating world models unifying the
cosmological dark sector (dark matter and dark energy). All the
models are described by a simplified version of the Chaplygin gas
Quartessence cosmology. It is found that even for $\Omega_k \neq 0$,
this Quartessence scenario depends only on a pair of parameters
which can severely be constrained by the cosmological tests. As an
example we perform a joint analysis involving the latest SNe type Ia
data and the recent Sloan Digital Sky Survey measurement of baryon
acoustic oscillations. In our analysis we have considered the SNe
type Ia Union sample compiled by Kowalski et al. (2008). At 95.4\%
(c.l.), we find for BAO + Union sample,
$\alpha=0.81^{+0.04}_{-0.04}$ and $\Omega_{\rm
Q4}=1.15^{+0.16}_{-0.17}$. The best fit for this simplified
Quartessence scenario is a spatially closed Universe and its reduced
$\chi^2$ is exactly the same of the flat concordance model
($\Lambda$CDM).
\end{abstract}

\pacs{98.80.Es; 95.35.+d; 98.62.Sb}
\maketitle

\section{Introduction}

The most plausible picture for the observed Universe seems to be
represented by a nearly flat scenario dominated by cold dark matter
(CDM) and a relativistic component endowed with large negative
pressure, usually named dark energy \cite{obs,Riess04,Kowalski08}. Although
having different status from a theoretical and observational
viewpoints, the actual nature of these dominant components remains
unknown until the present.  Therefore, in certain sense, one may say
that the modern general relativistic cosmology is plagued with the
so-called cosmological ``dark sector problem".

Recently, many cosmological models driven by dark matter and dark
energy have been proposed in the literature aiming to explain the
late time cosmic acceleration and other complementary observational
results\cite{rev123,lambdat,xmatter,sfield,DarkFluid,kamen,bilic,bento}.
Among these scenarios, a very interesting one was suggested by
Kamenshchik et al. \cite{kamen} and further developed by Bili\'c et
al. \cite{bilic} and Bento et al. \cite{bento}. It corresponds to a
class of world models dominated by an exotic fluid, named Chaplygin
gas (C-gas), which can be macroscopically characterized  by the
equation of state (EoS)
\begin{equation}\label{eq1}
p_C = -A/\rho_C^{\alpha},
\end{equation}
where $\alpha$ = 1 and $A$ is a positive constant related to the
present-day Chaplygin adiabatic sound speed, $v^2_s = \alpha
A/\rho_{C_{o}}^{1 + \alpha}$ ($\rho_{C_{o}}$ stands for the
current C-gas density).

The above equation for $\alpha \neq 1$ constitutes a generalization
of the original C-gas EoS proposed by Bento et al. in Ref.
\cite{bento}. One of its fundamental features comes from the fact
that the C-gas becomes pressureless at high redshifts, which
suggests a possible unification scheme for the cosmological ``dark
sector" (CDM plus dark energy). Scenarios driven by a C-gas (without
an extra CDM component) are usually termed quartessence models and
have  been largely explored in the literature \cite{quartessence}.

In most of these quartessence analyses, besides the present value of
the C-gas density parameter ($\Omega_C$), the above barotropic EoS
implies that one needs to constrain two additional free parameters,
namely, $A$ and $\alpha$ since the baryonic density ($\Omega_b$) may
be fixed a priori by using, for instance, nucleosynthesis or the
recent Cosmic Microwave Background (CMB) observations
\cite{Steigmann}. Therefore, in the context of a general
Friedman-Robertson-Walker (FRW) cosmologies quartessence scenarios
require at least 3 parameters to be constrained by the data (see for
instance, Bertolami et al. \cite{bert04}). In other words, there are
so many parameters to be constrained by the data, that a high degree
of degeneracy on the parametric space becomes inevitable. The common
solution in the literature to reduce the number of free parameters
(motivated by current CMB results) is to assume a flat geometry,
i.e., $\Omega_{\rm{Q4}} = 1 - \Omega_b$, where $\Omega_b$ and
$\Omega_{\rm{Q4}}$ stand, respectively, for the  baryons and C-gas
density parameters.

In the last few years, some generalizations of the  original C-gas
\cite{ZhWuZh06,ChLa05,MaHa05,GuZh05}, or even of its extended
version \cite{SeSc05} have appeared in literature. In these cases,
the number of free parameters is usually increased, and, as
consequence, the models become mathematically richer although much
less predictive from a physical viewpoint.

In a recent paper (from now on paper I) we took the opposite way,
that is,  we have proposed a large set of cosmologies driven by dark
energy plus a CDM component where the dark energy component was
represented by a simplified version of the Chaplygin gas whose
equation of state is described by just one free
parameter\cite{LCA106}. By adding the flatness condition the
resulting cosmology depends only of two free dimensionless
quantities, namely: the density  parameter, $\Omega_m$, and the
equation of state parameter, $\alpha$.

In this work we explore the results of paper I now for a
Quartessence version thereby reducing still more the parameter
space.  In this way, we discuss what we believe to be the simplest
Quartessence scenario, that is, the one with the smallest number of
free parameters. As we shall see, by an additional physical
condition, the allowed range of the $\alpha$ free parameter is also
restricted a priori, which makes not only the relevant parametric
space bidimensional - even for nonflat spatial sections - but also
(and more important) the model can be more easily discarded or
confirmed by the present set of observations since the range of its
free parameter is physically limited from causality considerations.
We test the viability of this simplified Quartessence approach by
discussing the constraints imposed from current SNe Ia observations,
compilation obtained by Supernovae Cosmology Project (SCP) group,
and Large Scale Structure (LSS) data. As we shall see the model
passes the background tests discussed here and its reduced
$\chi^{2}$ test is slightly smaller than the one of the $\Lambda
CDM$ model.

\section{A simplified Quartessence scenario}

Let us now consider that the geometrical properties of the observed
Universe are described by the general FRW line element
\begin{equation}
ds^2=dt^2-a(t)^2\left(\frac{dr^2}{1-kr^2} +r^2d\Sigma^2\right),
\end{equation}
where $a(t)$ is the scale factor, $d\Sigma^2$ is the area element on
the unit 2-sphere, $k=0,\pm1$ is the curvature parameter and we have
adopted the metric signature convention ($+$,$-$,$-$,$-$).
Throughout this paper we adopt units such that $c=1$. The matter
content of the Universe is assumed to be composed of a baryonic
component plus the quartessence C-gas fluid.

Since each component is separately conserved, one may integrate
out the energy conservation for the C-gas, $\dot{\rho}_{C} = -3H
(\rho_{C} + p_C)$, to obtain the following expression for its energy density
\cite{bento,quartessence,CAL04}
\begin{equation}\label{eq3}
\rho_{C} = \rho_{C_{o}}\left[A_s + (1 - A_s)a^{3(1 +
\alpha)}\right]^{\frac{1}{1 + \alpha}},
\end{equation}
where $A_s = A/\rho_{C_{o}}^{1 + \alpha}$ is a convenient dimensionless constant
(as usual, the subscript ``0" denotes present-day quantities). In the background
defined by (2), the Friedmann equation for a conserved C-gas plus the baryonic component reads
\begin{eqnarray}\label{eq4}
{\cal{H}} = \left[\Omega_{\rm{b}}(\frac{a_0}{a})^{3} +
\Omega_{\rm{Q4}}f(A_s, \alpha) + \Omega_{k}(\frac{a_0}{a})^{2}\right]^{1/2},
\end{eqnarray}
where ${\cal{H}} \equiv H/H_0$ ($H$ is the Hubble parameter), the
function $f(A_s, \alpha)$ is given by $f(A_s, \alpha)= [A_s + (1 -
A_s) (\frac{a_0}{a})^{3(\alpha + 1)}]^{\frac{1}{\alpha + 1}}$ and
$\Omega_k$ is the fractional contribution of the spatial curvature
to ${\cal{H}}$. Note that, besides the Hubble parameter $H_0$, we
still have 3 additional parameters in this case ($\alpha, A_s,
\Omega_{\rm{Q4}}$), since the baryonic contribution is defined to be
$\simeq 4.6\%$ from current CMB experiments
\cite{2008arXiv0803.0547K}. This is the standard treatment.
Therefore, the important aspect to be discussed at this point is how
to reduce the quartessence C-gas parameters based on reasonable
physical requirements?

In order to answer the above question, we follow the arguments of
Ref. \cite{LCA106}. Note that the C-gas adiabatic sound speed reads
\begin{equation}\label{eq5}
v_s^{2} = \frac{dp}{d\rho}= \alpha A/\rho_{C}^{1 + \alpha},
\end{equation}
which must be positive definite for a well-behaved gas (zero in the limit case
of dust). Note also that the present-day C-gas adiabatic sound speed is
$v_{so}^{2} = \alpha A/\rho_{C_{o}^{1 + \alpha}}$, or still
\begin{equation}\label{eq6}
v_{so}^{2} = \alpha A/\rho_{C_o}^{1 + \alpha} = \alpha A_s.
\end{equation}
Therefore, from the above equation one clearly see that if the parameter $A_s$ is a
function of the index $\alpha$, i.e., $A_s\rightarrow A_s(\alpha)$, the
number of free parameters is naturally reduced, and, as an extra bonus, the
positiviness of $v_s^{2}$ at any time, as well as its thermodynamic stability,
is naturally guaranteed. Clearly, among many possibilities the simplest choice
is $A_s \propto \alpha$, which we assume in this paper. In this case, $v_{so}^{2} =
\alpha^{2}$, or more generally, $v_s^{2} = {\alpha}^{2}(\rho_{Co}/\rho)^{\alpha}$.
Note also that, since the light speed is a natural cutoff for the sound propagation, it
follows that $v_{so}=|\alpha|\leq 1$, thereby restricting $\alpha$ to the interval [-1,1].
An additional constraint can still be imposed to this parameter. In fact, with
$A_s \propto \alpha$, the simplified C-gas EoS (1) becomes
\begin{equation}\label{eq7}
p_C = -\alpha
\rho_{Co}\left(\frac{\rho_{Co}}{\rho_{C}}\right)^{\alpha},
\end{equation}
so that a negative pressure is obtained only for positive values of $\alpha$. In other words,
this accounts to saying that the combined requirements from causality along with the observed
accelerating stage of the Universe naturally restrict the  parameter $\alpha$ to the interval
$0 < \alpha \leq 1$ \cite{jailson}.

\begin{figure}[t]
\centerline{\psfig{figure=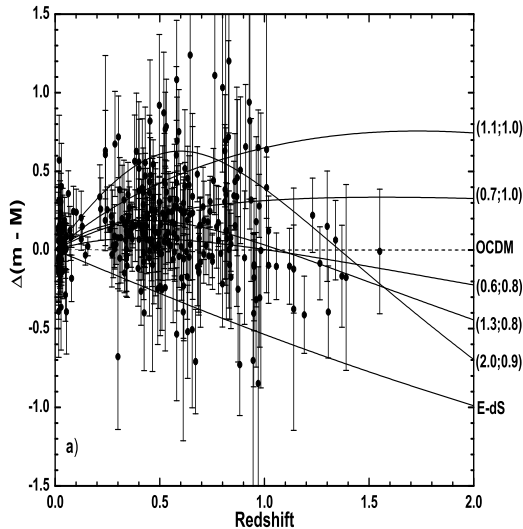,width=3.0truein,height=3.2truein}
\hskip 0.1in} \caption{We display the residual magnitudes with
respect to an empty ($\Omega_{total}=0$) universe for the SNe
samples Kowalski et al. (2008) \cite{Kowalski08}. The solid
curves are the predictions for the simplified quartessence scenario
characterized by the pair of parameters ($\Omega_{\rm Q4},
\alpha$).}
\end{figure}

Note that the simplified quartessence component preserves the unifying character
of the original C-gas, i.e., it behaves as a pressureless fluid (non-relativistic
matter) at high-$z$ while, at late times,  it approaches the quintessence behavior,
which now is fully characterized by the $\alpha$ parameter. However, note also
that, even in this limiting case, the sound speed is positive.

In this simplified approach, Eq.(\ref{eq4}) is rewritten as
\begin{eqnarray}\label{eq8}
{\cal{H}} = \left[\Omega_{\rm{b}}(\frac{a_0}{a})^{3} + \Omega_{\rm{Q4}}g(\alpha) +
\Omega_{k}(\frac{a_0}{a})^{2}\right]^{1/2},
\end{eqnarray}
where the function $g(\alpha)$ is simply given by $g(\alpha)=
[\alpha + (1 - \alpha)(\frac{a_0}{a})^{3(\alpha + 1)}]^{\frac{1}{\alpha + 1}}$,
so that the only remaining parameters to be determined in this unified
dark matter/energy scenario are $\alpha$ and $\Omega_{\rm{Q4}}$.
In what follows, we confront this simplified quartessence scenario
with some the most recent SNe Ia and Large Scale Structure (LSS) data.

\section{Observational Constraints}

\subsection{SNe Ia}

Let us first investigate the bounds arising from SNe Ia observations
on the SC-gas scenario described above. To this end we use the most
recent SNe Ia observations, namely, the Union compilation
\cite{Kowalski08}. It includes 13 independent sets with SNe
from the SCP, High-z Supernovae Search (HZSNS) team, Supernova
Legacy Survey and ESSENCE Survey, the older datasets, as well as the
recently extended dataset of distant supernovae observed with HST.
After selection cuts, the robust compilation obtained is composed by
307 SNe Ia events distributed over the redshift interval $0.015 \leq
z \leq 1.55$. Figure 1 shows residual magnitude versus redshift for
307 SNe type Ia from SCP Union compilation. The Union sample is
illustrated on a residual Hubble Diagram with respect to the empty
universe model ($\Omega _{total}=0$).

\begin{figure*}[t]
\centerline{\psfig{figure=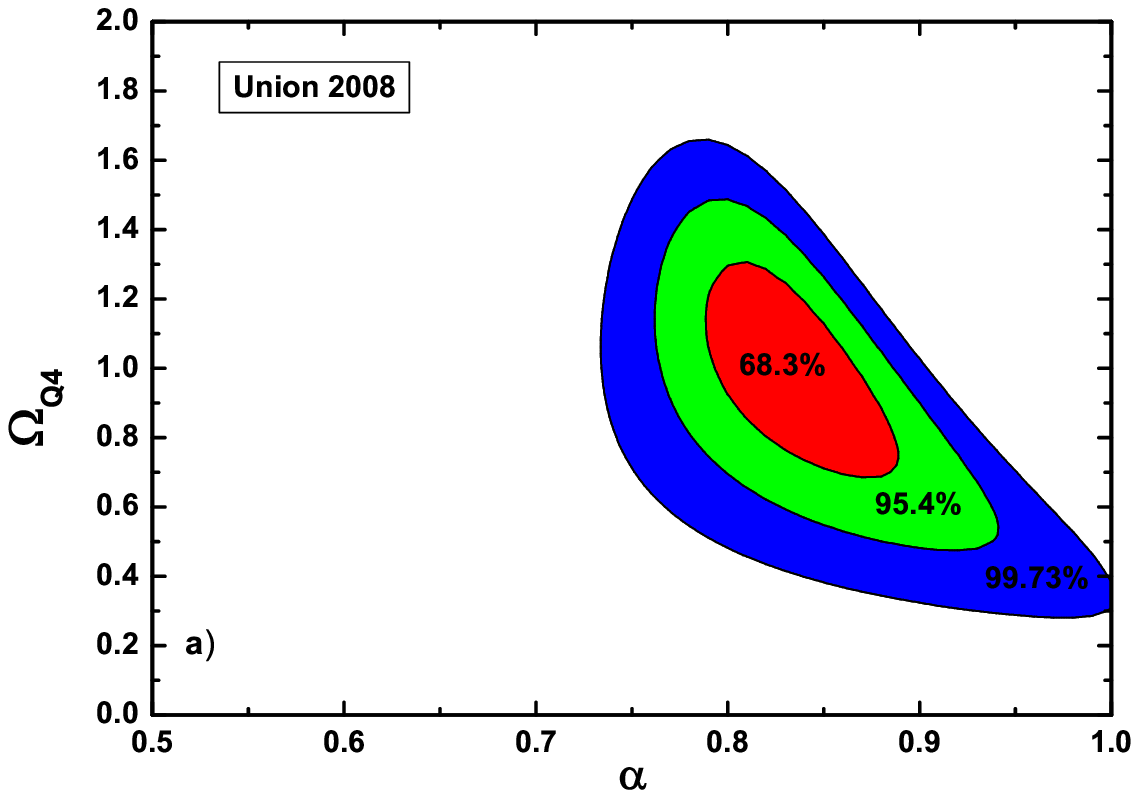,width=2.4truein,height=2.6truein}
\psfig{figure=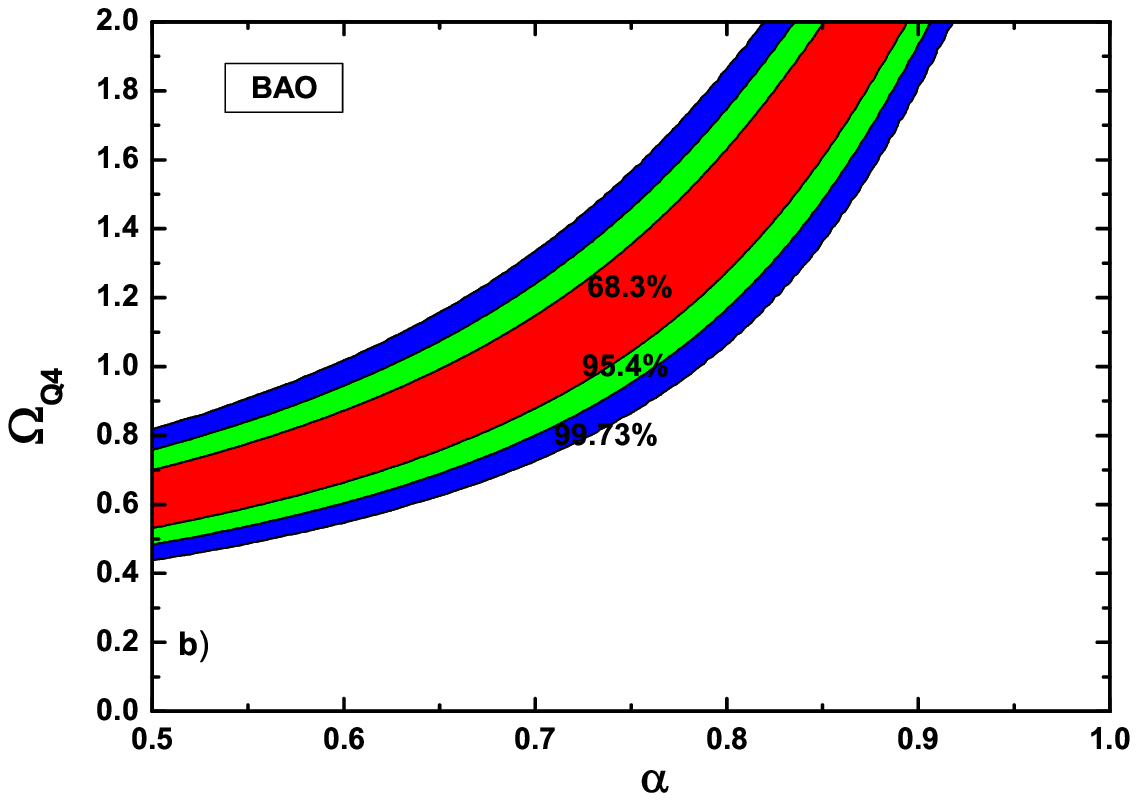,width=2.4truein,height=2.6truein}
\psfig{figure=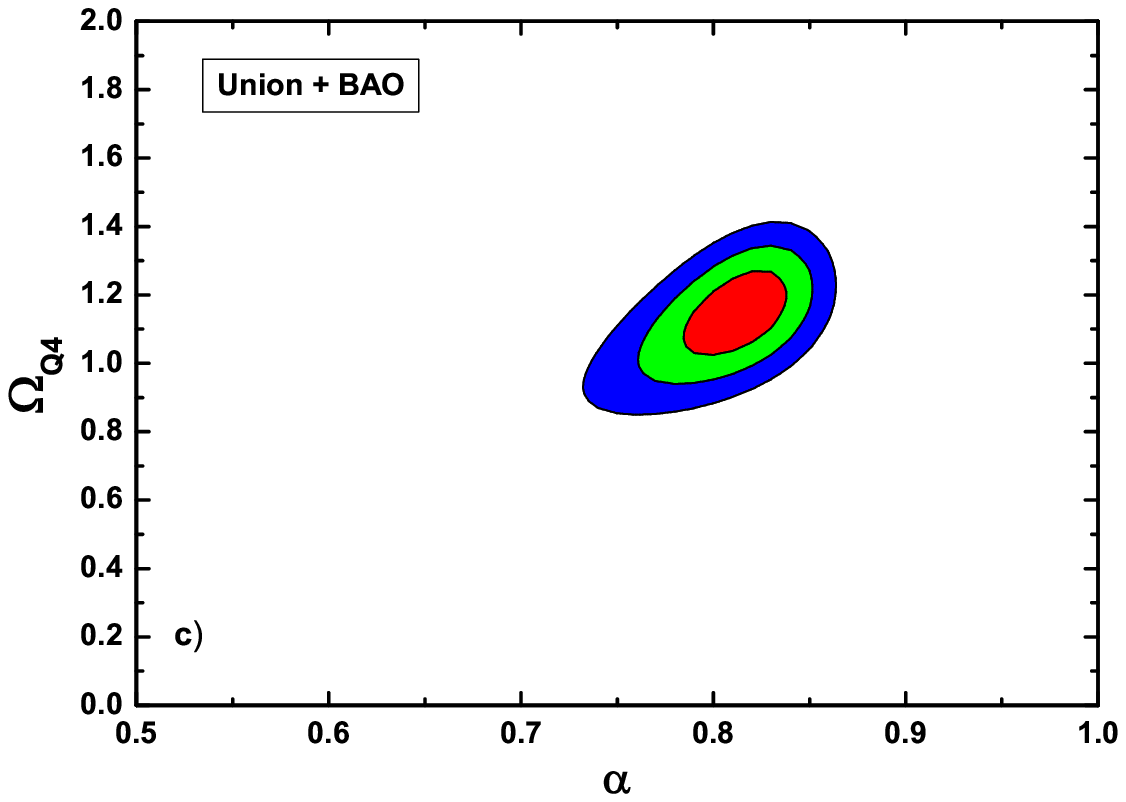,width=2.4truein,height=2.6truein} \hskip
0.1in} \vspace{-0.6cm} \caption{(a) Contours represent confidence
regions for Union data \cite{Kowalski08}. Our analysis
furnishes the regions on $\alpha=0.83^{+0.09}_{-0.06}$ and
$\Omega_{\rm Q4}=1.02^{+0.38}_{-0.45}$ ($2\sigma$). (b) Contours on
the space parameter ($\Omega_{\rm Q4}, \alpha$) from a joint
analysis involving the Sloan Sky Digital Survey (SDSS) baryon
acoustic oscillations. The corresponding $68.3\%$, $95.4\%$ and
$99.73\%$ c.l. are shown. (c) We display the results for the SNe
sample plus BAO. The best fit and confidence regions are
$\alpha=0.81^{+0.04}_{-0.04}$ and $\Omega_{\rm
Q4}=1.15^{+0.16}_{-0.17}$ ($2\sigma$).}
\end{figure*}

The predicted distance modulus for a supernova at redshift $z$, given a set of parameters $\mathbf{p}$, is
\begin{equation} \label{dm}
\mu_p(z|\mathbf{p}) = m - M = 5\mbox{log} d_L + 25,
\end{equation}
where $m$ and $M$ are, respectively, the apparent and absolute magnitudes,
the complete set of parameters is $\mathbf{p} \equiv (H_o, \Omega_{\rm{Q4}},
\alpha)$ and $d_L$ stands for the luminosity distance (in units of megaparsecs),
\begin{equation}
d_L = H_o^{-1}(1 + z)\frac{1}{\sqrt|\Omega_k|} {\rm \xi}
\left(\sqrt{|\Omega_k|}\int_{x'}^{1} {dx \over
x^{2}{\cal{H}}(x;\mathbf{p})}\right),
\end{equation}
with $x' = (1 + z)^{-1}$, ${\cal{H}}(x; \mathbf{p})$ the expression given by Eq.
(\ref{eq8}), and the function ${\rm \xi}(x)$ is defined as ${\rm \xi}(x) = \sin(x)$
for a closed universe, ${\rm \xi}(x) = \sinh(x)$ for an open universe and
${\rm \xi}(x) = x$ for a flat universe.

We estimated the best fit to the set of parameters $\mathbf{p}$ by using a
$\chi^{2}$ statistics
\begin{equation}
\chi^{2} = \sum_{i=1}^{N}{\frac{\left[\mu_p^{i}(z|\mathbf{p}) -
\mu_o^{i}(z|\mathbf{p})\right]^{2}}{\sigma_i^{2}}},
\end{equation}
with the parameters $\Omega_{\rm{Q4}}$ and $\alpha$ spanning the
interval [0,1] in steps of 0.01. In the above expression, $N = 307$,
$\mu_p^{i}(z|\mathbf{p})$ is given by Eq. (\ref{dm}),
$\mu_o^{i}(z|\mathbf{p})$ is the distance modulus for a given SNe Ia
at $z_i$, and $\sigma_i$ is the uncertainty in the individual
distance modulus. In our analysis, $H_0$ is considered a
\emph{nuisance} parameter so that we marginalize over it.

In Figures (2a) we plot the results of our statistical analysis.
Contours of constant likelihood (99.73$\%$, 95.4$\%$ and 68.3$\%$)
are shown in the parametric space $\alpha-\Omega_{\rm{Q4}}$. It
displays the results for the Union SCP compilation. Note that
although degenerate in $\Omega_{\rm Q4}$, the parameter $\alpha$ is
now considerably more restricted than in the standard C-gas approach
(see, e.g., Fig. 4 of Ref. \cite{jailson}). In particular, note also
that for any value of the C-gas density parameter, models with
$\alpha \lesssim 0.73$ are ruled out at 99.73\% level. The best-fit
model for this analysis occurs for $\Omega_{\rm{Q4}} = 1.02$ and
$\alpha = 0.83$ with $\chi^{2}_{\rm{min}} = 310.4$
($\chi^{2}_{\rm{min}}/\nu = 1.01$, where $\nu \equiv$ degrees of
freedom). At 95.4\% c.l. we also find $0.57 \leq \Omega_{\rm Q4}
\leq 1.40$ and $0.77 \leq \alpha \leq 0.92$.

\subsection{SNe Ia + LSS analysis}

In order to break possible degeneracies in the $\Omega_{\rm{Q4}} -
\alpha$ space, we study now the the joint constraints on this plane
from SNe Ia and LSS data. For the LSS data, we use the recent
measurements of the BAO peak in the large scale correlation function
detected by Eisenstein et al. \cite{BAO} using a large sample of
luminous red galaxies from the SDSS Main Sample. The SDSS BAO
measurement provides ${\cal{A}} = 0.469(n_S/0.98)^{-0.35} \pm
0.017$, with ${\cal{A}}$ defined as
\begin{eqnarray}
{\cal{A}} \equiv {\Omega_{\rm{M}}^{1/2} \over
{{\cal{H}}(z_{\rm{BAO}};\mathbf{p})}^{1/3}}[\frac{1}{z_{\rm{BAO}}
\sqrt{|\Omega_k|}}
\\ {\rm{\xi}} \left(\sqrt{|\Omega_k|}\Gamma(z_{\rm{BAO}};\mathbf{p})\right)]^{2/3}, \nonumber
\end{eqnarray}
where $z_{\rm{BAO}} = 0.35$, ${\cal{H}}(z_{\rm{BAO}};\mathbf{p})$ is
given by Eq. (8), and we take the scalar spectral index $n_S =
0.96$, as given by Komatsu et al. \cite{2008arXiv0803.0547K}. In the
above expression, $\Gamma(z_{\rm{BAO}})$ is the dimensionless
comoving distance to $z_{\rm{BAO}}$, and $\Omega_M = \Omega_b + (1 -
\alpha)\Omega_{Q4}$, where $\Omega_b$ is the baryonic component and
$(1 - \alpha)\Omega_{Q4}$ is the portion of the Chaplygin gas that
acts like dark matter. It should be noticed that the dark matter
contribution was derived here by using the separation proposed in
Ref. \cite{bento04} (see also \cite{Makler}.)

As shown in Panel (2b) the regions representing the constraints from
SDSS BAO measurements on the parameter space $\Omega_{\rm{Q4}} -
\alpha$ are approximately orthogonal to those arising from SNe Ia
data, which indicates that possible degeneracies in this plane may
be broken by this combination of observational data. Figure (2c)
shows the results of our joint analyses for the BAO + Union sample.
Note that the available parametric plane in both cases are
considerably reduced relative to the former analyses (Figs. 2a, 2b
and 2c). For the BAO + Union sample we find
$\alpha=0.81^{+0.04}_{-0.04}$ and $\Omega_{\rm
Q4}=1.15^{+0.16}_{-0.17}$ at 95.4\% (c.l.) with $\Omega_k =
-0.19^{+0.17}_{-0.16}$. This best-fit scenario corresponds to a
closed accelerating universe with $q_0 \simeq -0.7$, a total age of
the Universe of $t_o \simeq 10h^{-1}$ Gyr, and a D/A transition
redshift (from deceleration to acceleration) $z_{\rm{D/A}} \simeq
0.5$.

At this point, it is interesting to compare the above constraints
with some independent  analyses. In the context of the $\Lambda$CDM
model, for instance, the WMAP 5y constrains the curvature parameter
to be $\Omega_k= -0.099^{+0.085}_{-0.100}$ ($95$\%)
\cite{2008arXiv0803.0547K} in nice agreement with our result.  In
the same vein, the age of the Universe falls on the interval
13.5-14.0 Gyr, or equivalently, $9.3 h^{-1} < t_o < 10.5 h^{-1}$ Gyr
which is also comparable  with the above values. In addition, recent
kinematic studies (with no gravity theory) using SNe type Ia also
leads to the constraints $-0.5\lesssim q_0\lesssim -1$ and
$0.3\lesssim z_{\rm{D/A}}\lesssim 0.9$ ($1\sigma$)
\cite{CunhaLima08}. Finally, it is also worth noticing that many
alternative scenarios unifying dark matter and dark energy and even
accelerating cosmologies with no dark energy have  been proposed in
the literature \cite{DarkFluid,CDMAC}. Usually, such models are able
to explain not only the present accelerating expansion, but also the
majority of the so-called background tests (see the paper by Ellis
et al. \cite{DarkFluid} for a general analysis involving different
scenarios).

\section{Final remarks}

A considerable amount of observational evidence suggests that the
current evolution of our Universe is fully dominated by two dark
components, the so-called dark matter and dark energy. The nature
of these components, however, is a tantalizing mystery at present,
and it is not even known if they constitute two separate substances.
In this paper, we have argued that one of the candidates for a
unifying dark matter/dark energy scenario, a C-gas quartessence
whose EoS is given by Eq. (1), may have a very simplified
description. We have postulated that if $A_s$ is a function
of the index $\alpha$ the resulting FRW cosmology (with
arbitrary curvature) can be completely described only by
a pair of parameters ($\alpha$, $\Omega_{\rm Q4}$). For
the sake of simplicity, we have considered $A_s \propto \alpha^n$ with $n = 1$.

By considering this class of parameterization we have investigated
the constraints from current SNe Ia and LSS data. We have shown
that, differently from the original C-gas models (in which the value
of the index $\alpha$ is completely degenerated) a joint analysis
involving these data sets restricts considerably the
$\Omega_{\rm{Q4}} - \alpha$ parametric space [Fig. (2c)] with
$\alpha=0.81^{+0.04}_{-0.04}$ and $\Omega_{\rm
Q4}=1.15^{+0.16}_{-0.17}$. At the level of SNe Ia data and BAO, we
may conclude that this class of quartessence scenario passes this
combination of tests, thereby providing an interesting possibility
to a dark matter/dark energy unification. It is worth noticing that
the best-fit for this simplified quartessence scenario corresponds
to a spatially closed universe and, with the same number of
parameters, the $\chi^{2}_{min}=311.1$ is slightly smaller than the
one of the flat concordance model ($\Lambda$CDM). Still more
important, the reduced $\chi^{2}$ (by degree of freedom) in our
curved scenario, $\chi^{2}/dof = 1.02$, is exactly the same of the
cosmic concordance  flat model.

\begin{acknowledgments}
JASL and JSA are partially supported by the Conselho Nacional de
Desenvolvimento Cient\'{\i}fico e Tecnol\'{o}gico (CNPq - Brazil).
JVC and JASL are also supported by Funda\c{c}\~{a}o de Amparo \`a Pesquisa do
Estado de S\~ao Paulo (FAPESP), 2005/02809-5 and  04/13668-0, respectively.
\end{acknowledgments}

\end{document}